\documentclass[journal]{IEEEtran}
\usepackage{times}
\usepackage{stfloats}
\usepackage{graphicx}
\usepackage{amsmath}
\usepackage{amssymb}
\usepackage{amsthm}
\usepackage{mathtools,algorithm}
\usepackage{bm}
\usepackage{xcolor,colortbl}
\usepackage{multirow}
\usepackage[T1]{fontenc}
\usepackage{algpseudocode}
\usepackage{booktabs} 
\usepackage[section]{placeins} 
\usepackage{array}
\usepackage{varwidth}
\usepackage{listings}
\usepackage{gensymb}
\usepackage{url}
\usepackage{subcaption} 

\usepackage{pgfplots}
\usepackage{pgfplotstable}
\pgfplotsset{compat=1.18}

\usepackage[breaklinks=true,colorlinks,bookmarks=false,citecolor=blue]{hyperref}

\begin{document}

\title{Leveraging Mathematical Reasoning of LLMs for Efficient GPU Thread Mapping}

\author{Jose Maureira, Crist\'obal A. Navarro, Hector Ferrada, and Luis Veas-Castillo%
\thanks{Jose Maureira, Crist\'obal A. Navarro, Hector Ferrada, and Luis Veas-Castillo are with the Instituto de Informática, Universidad Austral de Chile, Valdivia, Chile, and the Temporal research group (\href{http://temporal.uach.cl}{http://temporal.uach.cl}). Corresponding author: Jose Maureira (e-mail: jose.maureira01@alumnos.uach.cl).}
}

\maketitle

\begin{abstract}
Mapping parallel threads onto non-box-shaped domains is a known challenge in GPU computing; efficient mapping prevents performance penalties from unnecessary resource allocation. Currently, achieving this requires significant analytical human effort to manually derive bespoke mapping functions for each geometry. This work introduces a novel approach leveraging the symbolic reasoning of Large Language Models (LLMs) to automate this derivation entirely through in-context learning. Focusing on state-of-the-art open-weights models, we conducted a rigorous comparative analysis across spatial domains of increasing complexity. Our results demonstrate that modern local LLMs successfully infer exact $O(1)$ and $O(\log N)$ mapping equations for complex 2D/3D dense domains and 2D fractals, vastly outperforming traditional symbolic regression methods. Crucially, we profile the energetic viability of this approach on high-performance infrastructure, distinguishing between the code-generation and execution phases. While one-time inference incurs a high energy penalty---particularly for reasoning-focused models like DeepSeek-R1---this is a single upfront investment. Once integrated, the generated analytical kernels eliminate block waste entirely, yielding massive energy and time savings (e.g., up to $4833\times$ speedup and $2890\times$ energy reduction) during actual GPU workloads. Finally, we identify a current ``reasoning ceiling'' when these models face highly recursive 3D fractals (e.g., the Menger Sponge). This limitation benchmarks the present maturity of open-weight architectures, charting a viable path toward fully automated, energy-efficient GPU resource optimization.
\end{abstract}


\IEEEpeerreviewmaketitle

\section{Introduction}
\label{sec:introduction}
The GPU has become an indispensable tool for accelerating a vast range of applications in science and engineering, driven by its massively parallel architecture \cite{ref1, ref2}. The central paradigm of GPU programming involves subdividing a problem into many parts and solving them with thousands of parallel threads, which are organized into hierarchical structures such as blocks and grids. Maximizing the performance of these powerful devices hinges on a fundamental principle: ensuring that every thread is engaged in useful computation, thereby minimizing idle resources.

\begin{figure}[ht!]
 \centering
 \includegraphics[width=\columnwidth]{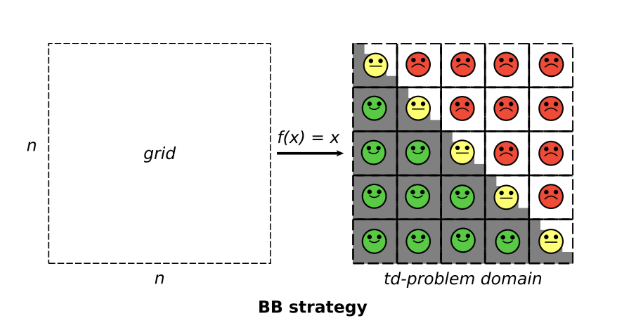}
 \caption{Illustration of how the classic BB mapping is not efficient for a 2D triangular domain. The red blocks represent wasted computational resources (threads/blocks) that are allocated but fall outside the useful problem area. (Adapted from \cite{ref1}).}
 \label{fig:bb_waste}
\end{figure}

\begin{figure}[ht!]
 \centering
 \includegraphics[width=\columnwidth]{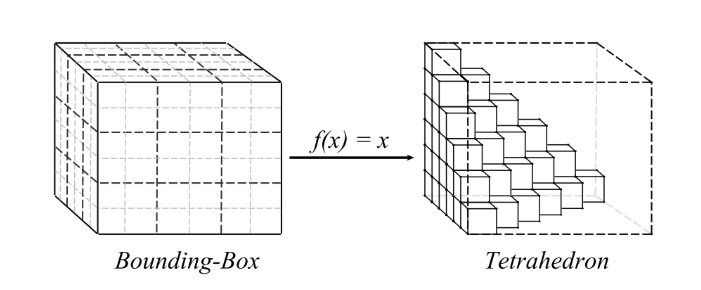}
 \caption{Conceptual representation of the inefficient BB mapping for a 3D triangular (tetrahedral) domain. A large portion of the allocated cube is wasted as it falls outside the problem space. (Adapted from \cite{ref2}).}
 \label{fig:tetra_waste}
\end{figure}

While standard grid-based thread mapping is effective for problems defined on regular domains (e.g., vectors, arrays, tables, matrices, boxes, etc), a significant number of scientific challenges involve complex, irregular geometries, leading to unpredictable control flow and memory access patterns \cite{ref26}. In fields like computational fluid dynamics, finite element analysis, and molecular simulations, the computational space is often non-uniform \cite{ref3, ref4}. For these domains, naive mapping strategies, such as using a simple bounding box (BB), are inefficient. As illustrated for 2D triangular domains (see Figure~\ref{fig:bb_waste}) and 3D tetrahedral domains (see Figure~\ref{fig:tetra_waste}), a large fraction of threads may be allocated outside the actual problem domain, just be discarded at runtime, thus wasting valuable computational cycles and energy.

Efficient thread mapping on GPUs is therefore essential for maximizing computational efficiency in various complex geometric domains. However, one of the main challenges lies in the high human effort required to manually formulate symbolic equations for each specific domain. Researching on this field had historically involved a high amount of analytical effort in order to come up with new equations \cite{ref5, ref6, ref7, ref8}, limiting the speed of progress for efficient GPU thread mapping for highly complex domains. 

This work takes a different route in GPU thread mapping research; it considers the recent advancements in open-weights large language models (LLMs) and how they can redefine the way research is done for this class of problems \cite{ref9, ref14, ref16, ref17}. Unlike traditional symbolic regression methods that function primarily as data-fitting algorithms, state-of-the-art LLMs with sophisticated reasoning capabilities can perform a higher-level task akin to algorithm induction or pattern reverse-engineering, entirely with local infrastructure. The main contributions of this work are:
\begin{itemize}
    \item We present a novel framework for automatically inferring GPU thread mapping equations for complex domains using in-context learning with local, open-source LLMs.
    \item We provide a comprehensive block-level performance and energy benchmark on high-performance infrastructure, revealing the critical trade-offs between model size, reasoning paradigms (e.g., Chain-of-Thought), and code execution efficiency.
    \item We identify a current ``reasoning ceiling'' for open-source models on highly recursive 3D fractals, establishing a benchmark for the present and upcoming open-weight models in automated mathematical derivation.
\end{itemize}

\section{Background and Related Work}
\label{sec:related_work}
This section situates our work within the established literature of GPU computing and the rapidly evolving field of symbolic regression.

\subsection{Efficient GPU Thread Mapping for Complex Domains}
The core problem is to define a bijective mapping $\lambda \in \mathrm{Z}_{+}^{k} \mapsto \mathrm{Z}_{+}^{d}$ from a k-dimensional set of thread coordinates in grid space to a set of d-dimensional coordinates in data domain space \cite{ref1}. This problem space relies on the fundamental GPU thread mapping principles described in canonical texts \cite{ref27}.

Previous research has systematically explored this problem, manually deriving mapping functions for various non-trivial domains, including triangular, tetrahedral, and fractal spaces \cite{ref1, ref2, ref3, ref5, ref6, ref7, ref8}. A canonical example from this body of work is the mapping function for a 2D lower triangular domain \cite{ref1}, which relates a linear index $\lambda$ to a coordinate pair $(i, j)$. The forward mapping function, $g(\lambda)$, is given by:
\begin{equation}
g(\lambda) = (i, j) = \left( \left\lfloor \sqrt{\frac{1}{4} + 2\lambda - \frac{1}{2}} \right\rfloor, \lambda - \frac{i(i+1)}{2} \right)
\end{equation}
This equation, while elegant, is non-trivial to derive from first principles. It involves an understanding of triangular numbers and requires a specific analytical insight to formulate. The manual effort required to produce such equations for progressively more complex domains escalates significantly. This work aims to automate the discovery of such functions.

\subsection{Symbolic Equation Inference using Deep Learning}
Symbolic regression (SR), the automated discovery of mathematical expressions from data, presents a promising avenue for such automation. The field has evolved from genetic programming to incorporate modern deep learning methods, including transformer-based architectures for sequence-to-sequence generation \cite{ref12}.

Models such as \textit{Neural Symbolic Regression that Scales} \cite{ref9} and \textit{SymFormer} \cite{ref13} utilize encoder-decoder structures to generate equation skeletons from data. Other approaches have integrated prior knowledge \cite{ref10} or used Monte-Carlo tree search \cite{ref11}. However, these methods are fundamentally designed as continuous data-fitting algorithms aimed at minimizing numerical error (e.g., mean squared error). This poses a critical theoretical limitation for GPU thread mapping, which operates strictly in the discrete integer domain. An approximation, no matter how numerically close, is inherently invalid for indexing array addresses or hardware threads, as absolute algorithmic precision is mandatory. Therefore, traditional SR networks are structurally unsuited for deriving exact mappings.

\subsection{Large Language Models in Symbolic Reasoning}
The recent and rapid advancements in LLMs have opened a new frontier for symbolic discovery \cite{ref14, ref15, ref16, ref17}. Recent comprehensive surveys highlight the transformative evolution of these models in automated code generation \cite{ref24} and mathematical reasoning \cite{ref25}, shifting their utility from pure natural language processing to complex algorithmic problem-solving. Instead of being trained specifically for continuous numerical SR, these models leverage their vast, pre-existing knowledge base to perform logical and symbolic reasoning tasks. A prominent method is \textbf{in-context learning}, where an LLM is guided to a solution through a prompt containing a few examples of the desired input-output behavior \cite{ref14, ref17}. This "few-shot" approach transforms the task from a continuous data-fitting problem into one of discrete pattern recognition and logical reverse-engineering, aligning perfectly with the goal of deriving exact, absolute mapping algorithms. This recent mathematical reasoning capability of modern LLMs is the ground from which this work builds upon. 

\begin{figure*}[ht!]
 \centering
 \includegraphics[width=1\textwidth]{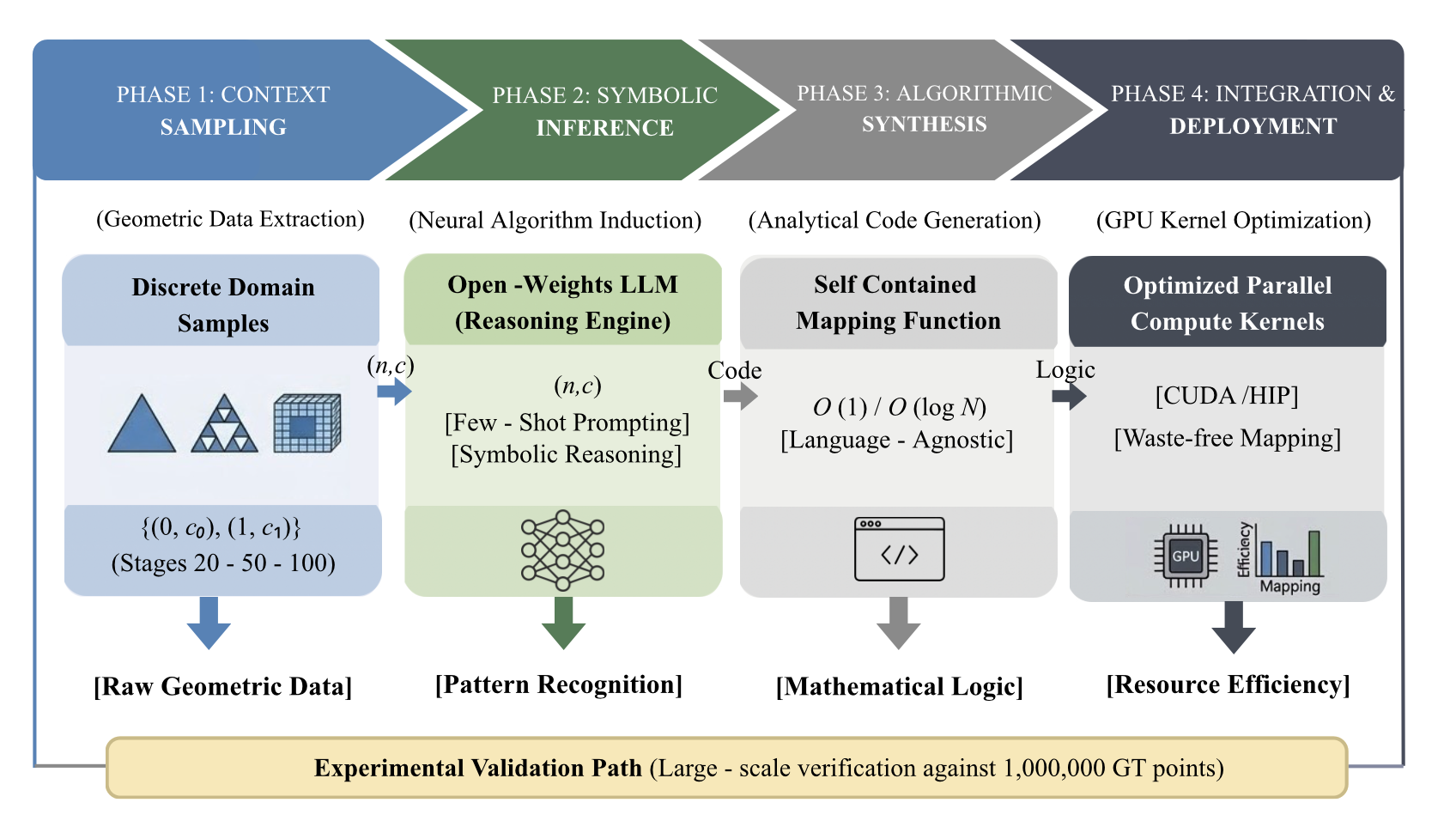}
 \caption{Overview of the proposed automated discovery pipeline. 
          (1) Data extraction from the target domain, (2) Neural symbolic reasoning for algorithm induction, 
          (3) Automated synthesis of the analytical mapping code, 
          and (4) Integration and deployment of the discovered logic.}
 \label{fig:pipeline_mother}
\end{figure*}

\section{Leveraging LLMs for more efficient GPU thread mapping}
\label{sec:llm_leverage}

The core of our proposal is to transform the manual, time-consuming analytical process of deriving mapping functions into an automated pipeline driven by the symbolic reasoning of open-weight LLMs. This approach treats geometric mapping not as a data-fitting problem, but as an algorithmic induction task where the model discovers the underlying map.

\subsection{Algorithmic Induction via In-Context Learning}
Unlike traditional symbolic regression, the use of LLMs allows for true "Algorithm Induction". By providing a sequence of coordinates as context, we trigger the model's ability to recognize complex patterns in integer sequences. This process is language-independent; the model effectively infers the abstract mathematical relationship required for absolute precision indexing and expresses it through functional code ready for parallel environments.

\subsection{Upfront Reasoning vs. Permanent Savings}
The procedure consists of a one-time "Reasoning Phase" where the model expends computational effort to derive the symbolic solution. Once the analytical logic is identified, it represents a permanent architectural optimization. This shifts the burden from a naive, wasteful runtime check (Bounding Box) to a one-time intelligent derivation, leading to the massive energy savings discussed in Section~\ref{sec:results}.

\subsection{The Automated Workflow and Experimental Validation}
The workflow is divided into four distinct phases as shown in Fig.~\ref{fig:pipeline_mother}. It is important to distinguish between the \textit{operational framework} and the \textit{experimental validation} presented in this study:

\begin{enumerate}
    \item \textbf{Context Sampling:} Extraction of the first $N$ elements of the domain (Stage 20, 50, or 100) to define the pattern. These initial points are generated trivially via standard sequential CPU execution.
    \item \textbf{Symbolic Inference:} Neural induction of the mapping algorithm. The model is guided by a \textbf{few-shot structured prompt} that defines the operational constraints, the mathematical nature of the expected output, and provides the sampled coordinate context.
    \item \textbf{Algorithmic Synthesis:} Generation of a self-contained analytical code block ($O(1)$ or $O(\log N)$) that implements the discovered logic.
    \item \textbf{Integration and Deployment:} Direct application of the synthesized code into GPU kernels to optimize the amount of threads required.
\end{enumerate}

We emphasize that the large-scale validation (against $10^6$ points) and the empirical energy/time profiling described in the following sections are part of our \textit{experimental protocol} to formally prove the correctness and viability of the method. The value $10^6$ is strictly an evaluation parameter; the inferred analytical mapping functions naturally generalize to arbitrary problem sizes, bounded only by the hardware's numerical representation limits (e.g., 32-bit or 64-bit registers). In a production environment, once the model's reasoning reliability is established for a class of geometries, the discovery remains an autonomous process that does not require prior knowledge of the full ground truth dataset.

\section{Framework for Evaluating Symbolic Inference}
\label{sec:methodology}

This section details the experimental protocol designed to evaluate the accuracy and efficiency of the selected LLMs. To simulate a realistic scenario of algorithmic discovery from scarce data, we employed an in-context learning approach with varying levels of information.

\subsection{Experiment Design}
\begin{figure}[ht!]
 \centering
 \includegraphics[width=\columnwidth]{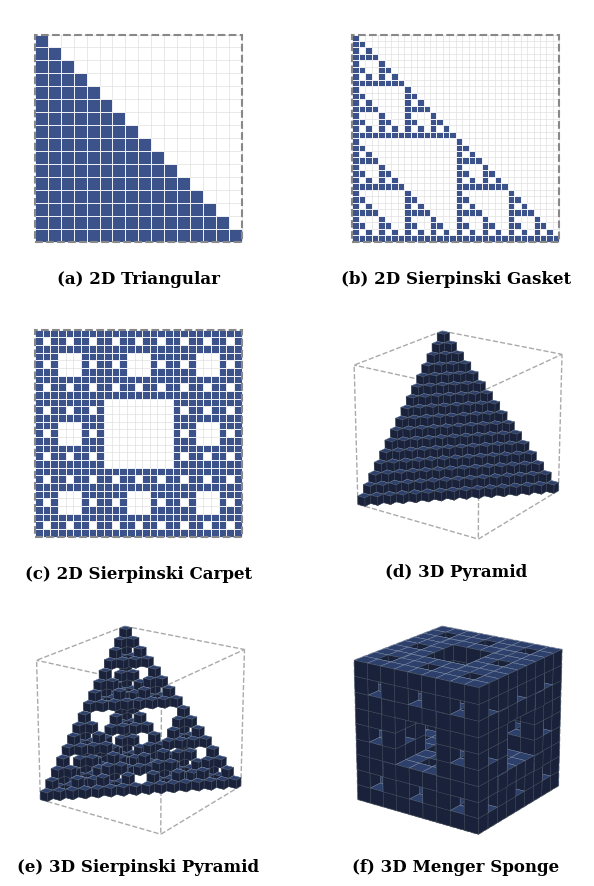}
 \caption{Visual overview of the six evaluated computational domains: 
          (a) 2D Triangular, (b) 2D Sierpinski Gasket, 
          (c) 2D Sierpinski Carpet, (d) 3D Pyramid, 
          (e) 3D Sierpinski Pyramid, and (f) 3D Menger Sponge. }
 \label{fig:domains_overview}
\end{figure}

We evaluated the models across six computational domains of increasing geometric and algorithmic complexity. Figure \ref{fig:domains_overview} visualizes these target geometries, and Table \ref{tab:domains_overview} summarizes the exact Ground Truth (GT) mathematical logic required to correctly map a linear thread index $\lambda$ to the spatial coordinates $\mathbf{c}$ in each domain. The analytical and fractal equations presented are extracted directly from the established literature on non-linear block-space mapping \cite{ref1, ref2, ref3, ref4, ref6, ref7, ref8}.

\begin{table*}[ht!]
\centering
\caption{Summary of evaluated computational domains, their geometric complexity, and the analytical or algorithmic Ground Truth equations required for exact thread mapping, along with their respective literature sources.}
\label{tab:domains_overview}
\renewcommand{\arraystretch}{1.8}
\resizebox{\textwidth}{!}{%
\begin{tabular}{llcp{11.5cm}}
\toprule
\textbf{Domain} & \textbf{Type} & \textbf{Complexity} & \textbf{Ground Truth Mapping Logic $\lambda \rightarrow \mathbf{c}$} \\ \midrule

\textbf{2D Triangular} \cite{ref1, ref4} & Dense & $O(1)$ & 
$x = \left\lfloor \sqrt{\frac{1}{4} + 2\lambda} - \frac{1}{2} \right\rfloor, \quad y = \lambda - \frac{x(x+1)}{2}$ \\

\textbf{3D Pyramid} \cite{ref2, ref4} & Dense & $O(1)$ & 
$v = \frac{\sqrt[3]{\sqrt{729\lambda^2 - 3} + 27\lambda}}{3^{2/3}} + \frac{1}{\sqrt[3]{3}\sqrt[3]{\sqrt{729\lambda^2 - 3} + 27\lambda}} - 1, \quad z = \lfloor v \rfloor, \quad (x,y) = f_{\text{2D\_Tri}}\left(\lambda - \frac{z(z+1)(z+2)}{6}\right)$ \\

\textbf{2D Sierpinski Gasket} \cite{ref3, ref6} & Fractal & $O(\log_3 N)$ & 
$(x,y) = \sum_{i=0}^{\lfloor \log_3 \lambda \rfloor} \vec{v}_{d_i} 2^i, \quad \text{where } \lambda = \sum d_i 3^i, \; \vec{v}_{d_i} \in \{(0,0), (1,0), (0,1)\}$ \\

\textbf{2D Sierpinski Carpet} \cite{ref6, ref7} & Fractal & $O(\log_8 N)$ & 
$(x,y) = \sum_{i=0}^{\lfloor \log_8 \lambda \rfloor} \vec{v}_{d_i} 3^i, \quad \text{where } \lambda = \sum d_i 8^i, \; \vec{v}_{d_i} \in \{0,1,2\}^2 \setminus \{(1,1)\}$ \\

\textbf{3D Sierpinski Pyramid} \cite{ref7} & Fractal & $O(\log_4 N)$ & 
$(x,y,z) = \sum_{i=0}^{\lfloor \log_4 \lambda \rfloor} \vec{v}_{d_i} 2^i, \quad \text{where } \lambda = \sum d_i 4^i, \; \vec{v}_{d_i} \in \{(0,0,0), (1,0,0), (0,1,0), (0,0,1)\}$ \\

\textbf{3D Menger Sponge} \cite{ref7} & Fractal & $O(\log_{20} N)$ & 
$(x,y,z) = \sum_{i=0}^{\lfloor \log_{20} \lambda \rfloor} \vec{v}_{d_i} 3^i, \quad \text{where } \lambda = \sum d_i 20^i, \; \vec{v}_{d_i} \in \{0,1,2\}^3 \setminus \mathcal{V}_{\text{voids}}$ \\ \bottomrule
\end{tabular}%
}
\end{table*}

As presented in Table \ref{tab:domains_overview}, dense domains rely on the exact mathematical inverse of arithmetic progressions and tetrahedral numbers (requiring floating-point square and cubic roots). In contrast, the mapping for fractal domains leverages a systematic base-$B$ digit decomposition, where the linear index $\lambda$ is decomposed into its constituent digits $d_i$, which in turn dictate the scaled spatial translation vectors $\vec{v}_{d_i}$ applied at each recursive depth. 

For each domain-model pair, we followed a strict evaluation pipeline designed to simulate algorithmic discovery from scarce data:

\subsubsection{In-Context Learning Stages}
The models were prompted with a sequence of integer coordinates representing the domain. We defined three "Stages" of information density to test the models' ability to generalize from small samples:
\begin{itemize}
    \item \textbf{Stage 20:} The prompt includes the first 20 points of the domain. This tests extreme few-shot inference.
    \item \textbf{Stage 50:} The prompt includes the first 50 points.
    \item \textbf{Stage 100:} The prompt includes the first 100 points, providing a richer context for pattern recognition.
\end{itemize}

\subsubsection{Validation against Ground Truth}
Once a model generates a Python function, we validate it against a massive Ground Truth (GT) dataset of $N=1,000,000$ points. This ensures that the inferred logic holds for the entire domain and not just the few examples provided in the prompt. Specifically, the validation verifies that the inferred function 
produces a \textit{bijective} mapping over a ground truth dataset of 
$N = 1{,}000{,}000$ points: every valid domain coordinate must be 
visited exactly once, with no repetitions and no omissions. To capture 
the full spectrum of model behavior, correctness is assessed under two 
complementary criteria. The first, \textit{Ordered}, checks whether 
the model's output matches the ground truth sequence index-by-index, 
rewarding exact algorithmic reproduction. The second, 
\textit{Any-order}, checks whether all unique domain coordinates are 
covered regardless of traversal order, identifying solutions that 
recover the correct geometry but with a permuted index sequence.

\subsubsection{Evaluation Metrics}
We classify the performance using two accuracy metrics and one efficiency metric:
\begin{enumerate}
    \item \textbf{Ordered:} The percentage of indices $n \in [0, 10^6]$ where the model's output $(x,y,z)$ exactly matches the GT vector sequence.
    \item \textbf{Any-order:} The percentage of unique GT coordinates found by the model, regardless of the index order. This identifies "Silver Standard" solutions where the geometric shape is correct, but the mapping order is permuted.
    \item \textbf{Big-O Efficiency:} We perform static and dynamic analysis to ensure the generated code is optimal, extracting empirical measurements on GPU execution time and energy bounds.
\end{enumerate}

\section{Results and Discussion}
\label{sec:results}
This section presents the findings from the experiments conducted on the local Patagón supercomputer infrastructure, focusing on the performance, efficiency, and limitations of the open-source LLM ecosystem.

\subsection{LLM Symbolic Inference Accuracy}
\label{subsec:llm_accuracy}

We evaluated 11 state-of-the-art open-weight LLMs (including models from the Llama 3.3/4 \cite{ref19}, Qwen3 \cite{ref20}, DeepSeek \cite{ref21}, Gemma3 \cite{ref22}, Mistral \cite{ref23}, and GPT-OSS families) across the six computational domains. These specific families were selected because they represent the current frontier of open-source capabilities, encompassing a diverse spectrum of underlying architectures---from traditional dense transformers to scalable Mixture-of-Experts (MoE) \cite{ref18}---and training paradigms, including standard instruction-tuning and Reinforcement Learning (RL) driven Chain-of-Thought (CoT). Furthermore, these models are strictly comparable as they represent the leading open-source competitors systematically evaluated against each other in recent holistic benchmarks for code generation, algorithmic logic, and mathematical reasoning \cite{ref28, ref29}. The results demonstrate a clear stratification in reasoning capabilities, demonstrating that local models can handle complex discrete symbolic tasks. 

In the accuracy tables (\ref{tab:2d_triangular}--\ref{tab:3d_menger}), model names are abbreviated to fit the column width and append their parameter sizes. The specific models evaluated in this benchmark correspond to: \textbf{R1:70b} (\texttt{deepseek-r1:70b}), \textbf{Gem3:12b} and \textbf{Gem3:27b} (\texttt{gemma3}), \textbf{OSS:20b} and \textbf{OSS:120b} (\texttt{gpt-oss}), \textbf{Lla3.3:70b} (\texttt{llama3.3}), \textbf{Lla4:16x17b} (\texttt{llama4}), \textbf{Mist-N:12b} (\texttt{mistral-nemo}), \textbf{Nemo:70b} (\texttt{nemotron:70b}), and \textbf{Qw3:32b} and \textbf{Qw3:235b} (\texttt{qwen3}). All models were executed locally in GGUF format using their default configuration parameters. Entries marked with \textbf{(NC)} indicate non-compiling or structurally invalid code.

\textbf{2D Domains (Triangular \& Fractals):} 
For the basic \textbf{2D Triangular mapping} (Table \ref{tab:2d_triangular}), top-tier models like \texttt{OSS:120b}, \texttt{OSS:20b}, and \texttt{R1:70b} achieved 100\% exact Ordered accuracy, demonstrating that basic arithmetic series are well within the grasp of open models. In the fractal domains, results diverged. For the \textbf{2D Sierpinski Gasket} (Table \ref{tab:2d_gasket}), only \texttt{OSS} variants successfully inferred the logic. Notably, for the highly complex \textbf{2D Sierpinski Carpet} (Table \ref{tab:2d_carpet}), which requires recursive modular arithmetic, \texttt{OSS:120b} (at 100 points) and \texttt{Qw3:235b} (at 20/50 points) achieved 100\% Ordered accuracy. This proves that massive open models can infer complex fractal logic that defeats smaller architectures.

\textbf{3D Domains (Triangular \& Pyramid):} 
Table \ref{tab:3d_triangular} reveals robust performance in 3D spaces. Models such as \texttt{OSS:120b}, \texttt{OSS:20b}, and the \texttt{Qw3} family consistently solved the \textbf{3D Triangular mapping} with 100\% accuracy. An interesting phenomenon was observed with \texttt{R1:70b}: at 20 points, it deduced the correct geometric set, achieving 82.70\% Any-order accuracy, but required more context to construct the precise index sequence, initially yielding 0.11\% Ordered accuracy. For the \textbf{3D Sierpinski Pyramid} (Table \ref{tab:3d_pyramid}), \texttt{OSS:120b} was the only model to reach 100\% perfect mapping.

\textbf{The "Menger Limit" (3D Menger Sponge):} 
The \textbf{3D Menger Sponge} (Table \ref{tab:3d_menger}) is the hardest domain and acts as a frontier in symbolic reasoning for existing open-source LLMs. While these models excel at 2D fractals and dense 3D spaces, the Menger Sponge requires tracking cubic coordinate systems and recursive 3D void removals. The open-source ecosystem currently achieves an Any-order score of $< 1\%$ on this domain. This establishes a clear ``reasoning boundary'' for state-of-the-art open weight models, making it a benchmark for the next generation of LLMs.


\begin{table}[ht!]
\centering
\caption{Results for 2D Triangular mapping.}
\label{tab:2d_triangular}
\resizebox{\columnwidth}{!}{%
\begin{tabular}{lcccccc}
\toprule
\multirow{2}{*}{\textbf{Model}} & \multicolumn{2}{c}{\textbf{20 pts}} & \multicolumn{2}{c}{\textbf{50 pts}} & \multicolumn{2}{c}{\textbf{100 pts}} \\ \cmidrule(lr){2-3} \cmidrule(lr){4-5} \cmidrule(lr){6-7}
 & \textbf{Ord.} & \textbf{Any} & \textbf{Ord.} & \textbf{Any} & \textbf{Ord.} & \textbf{Any} \\ \midrule
R1:70b & \textbf{100\%} & \textbf{100\%} & \textbf{100\%} & \textbf{100\%} & \textbf{100\%} & \textbf{100\%} \\
Gem3:12b & 0.00\% & 0.00\% & 0.00\% & 1.27\% & 0.00\% & 1.83\% \\
Gem3:27b & 0.00\% & 50.05\% & 0.00\% & 1.27\% & 0.00\% & 50.05\% \\
OSS:120b & \textbf{100\%} & \textbf{100\%} & \textbf{100\%} & \textbf{100\%} & \textbf{100\%} & \textbf{100\%} \\
OSS:20b & 0.00\% & 0.71\% & \textbf{100\%} & \textbf{100\%} & \textbf{100\%} & \textbf{100\%} \\
Lla3.3:70b & \textbf{100\%} & \textbf{100\%} & 0.00\% & 0.00\% & 0.00\% & 0.14\% \\
Lla4:16x17b & 0.00\% & 0.71\% & 0.00\% & 1.27\% & 0.00\% & 0.01\% \\
Mist-N:12b & 0.00\% & 0.71\% & 0.00\% & 1.27\% & 0.00\% & 1.69\% \\
Nemo:70b & 0.00\% & 0.00\% & 0.00\% & 0.14\% & \textbf{100\%} & \textbf{100\%} \\
Qw3:235b & \textbf{100\%} & \textbf{100\%} & 0.14\% & 0.14\% & 0.00\% (NC) & 0.00\% \\
Qw3:32b & \textbf{100\%} & \textbf{100\%} & \textbf{100\%} & \textbf{100\%} & \textbf{100\%} & \textbf{100\%} \\
\bottomrule
\end{tabular}
}
\end{table}

\begin{table}[ht!]
\centering
\caption{Results for 2D Sierpinski Gasket.}
\label{tab:2d_gasket}
\resizebox{\columnwidth}{!}{%
\begin{tabular}{lcccccc}
\toprule
\multirow{2}{*}{\textbf{Model}} & \multicolumn{2}{c}{\textbf{20 pts}} & \multicolumn{2}{c}{\textbf{50 pts}} & \multicolumn{2}{c}{\textbf{100 pts}} \\ \cmidrule(lr){2-3} \cmidrule(lr){4-5} \cmidrule(lr){6-7}
 & \textbf{Ord.} & \textbf{Any} & \textbf{Ord.} & \textbf{Any} & \textbf{Ord.} & \textbf{Any} \\ \midrule
R1:70b & 0.00\% & 8.10\% & 4.57\% & 21.30\% & 0.00\% & 1.52\% \\
Gem3:12b & 0.00\% & 1.03\% & 0.00\% & 1.55\% & 0.00\% & 0.69\% \\
Gem3:27b & 0.00\% & 1.03\% & 0.00\% & 5.22\% & 0.00\% & 5.22\% \\
OSS:120b & 0.00\% & 8.10\% & \textbf{100\%} & \textbf{100\%} & \textbf{100\%} & \textbf{100\%} \\
OSS:20b & \textbf{100\%} & \textbf{100\%} & 0.00\% (NC) & 0.00\% & \textbf{100\%} & \textbf{100\%} \\
Lla3.3:70b & 0.00\% & 7.96\% & 0.00\% & 1.17\% & 0.00\% & 3.19\% \\
Lla4:16x17b & 0.00\% & 0.34\% & 0.00\% & 0.00\% & 0.00\% & 0.01\% \\
Mist-N:12b & 0.00\% & 0.00\% & 0.00\% & 3.09\% & 0.00\% & 0.01\% \\
Nemo:70b & 0.00\% & 8.10\% & 0.00\% & 8.10\% & 0.00\% & 8.10\% \\
Qw3:235b & 0.00\% (NC) & 0.00\% & 0.00\% & 0.00\% & 0.00\% (NC) & 0.00\% \\
Qw3:32b & 0.00\% & 8.10\% & 0.00\% & 0.01\% & 0.00\% (NC) & 0.00\% \\
\bottomrule
\end{tabular}
}
\end{table}

\begin{table}[ht!]
\centering
\caption{Results for 2D Sierpinski Carpet.}
\label{tab:2d_carpet}
\resizebox{\columnwidth}{!}{%
\begin{tabular}{lcccccc}
\toprule
\multirow{2}{*}{\textbf{Model}} & \multicolumn{2}{c}{\textbf{20 pts}} & \multicolumn{2}{c}{\textbf{50 pts}} & \multicolumn{2}{c}{\textbf{100 pts}} \\ \cmidrule(lr){2-3} \cmidrule(lr){4-5} \cmidrule(lr){6-7}
 & \textbf{Ord.} & \textbf{Any} & \textbf{Ord.} & \textbf{Any} & \textbf{Ord.} & \textbf{Any} \\ \midrule
R1:70b & 0.00\% & 0.58\% & 0.00\% & 0.00\% & 0.00\% & 37.08\% \\
Gem3:12b & 0.00\% & 0.58\% & 0.00\% & 0.39\% & 0.00\% & 0.58\% \\
Gem3:27b & 0.00\% & 0.39\% & 0.00\% (NC) & 0.20\% & 0.00\% & 1.04\% \\
OSS:120b & 0.00\% & 0.58\% & 0.01\% & 1.04\% & \textbf{100\%} & \textbf{100\%} \\
OSS:20b & 0.00\% & 0.58\% & 0.00\% (NC) & 0.00\% & 0.00\% & 0.58\% \\
Lla3.3:70b & 0.00\% & 0.39\% & 0.00\% & 0.39\% & 0.00\% & 0.46\% \\
Lla4:16x17b & 0.00\% & 0.58\% & 0.00\% & 1.04\% & 0.00\% & 1.56\% \\
Mist-N:12b & 0.00\% & 0.39\% & 0.00\% & 1.04\% & 0.00\% & 1.30\% \\
Nemo:70b & 0.00\% & 0.00\% & 0.00\% & 0.58\% & 0.00\% & 0.10\% \\
Qw3:235b & \textbf{100\%} & \textbf{100\%} & \textbf{100\%} & \textbf{100\%} & 0.00\% (NC) & 0.00\% \\
Qw3:32b & 0.00\% & 0.00\% & 0.00\% & 0.03\% & 0.00\% & 0.58\% \\
\bottomrule
\end{tabular}
}
\end{table}

\begin{table}[ht!]
\centering
\caption{Results for 3D Triangular mapping.}
\label{tab:3d_triangular}
\resizebox{\columnwidth}{!}{%
\begin{tabular}{lcccccc}
\toprule
\multirow{2}{*}{\textbf{Model}} & \multicolumn{2}{c}{\textbf{20 pts}} & \multicolumn{2}{c}{\textbf{50 pts}} & \multicolumn{2}{c}{\textbf{100 pts}} \\ \cmidrule(lr){2-3} \cmidrule(lr){4-5} \cmidrule(lr){6-7}
 & \textbf{Ord.} & \textbf{Any} & \textbf{Ord.} & \textbf{Any} & \textbf{Ord.} & \textbf{Any} \\ \midrule
R1:70b & 0.11\% & 82.70\% & \textbf{100\%} & \textbf{100\%} & 0.00\% & 0.00\% \\
Gem3:12b & 0.00\% & 0.02\% & 0.00\% & 0.02\% & 0.00\% & 0.02\% \\
Gem3:27b & 0.00\% & 0.00\% & 0.00\% & 0.00\% & 0.00\% & 17.17\% \\
OSS:120b & \textbf{100\%} & \textbf{100\%} & \textbf{100\%} & \textbf{100\%} & \textbf{100\%} & \textbf{100\%} \\
OSS:20b & 0.00\% (NC) & 0.00\% & \textbf{100\%} & \textbf{100\%} & \textbf{100\%} & \textbf{100\%} \\
Lla3.3:70b & 0.00\% & 0.00\% & 0.00\% & 17.16\% & 0.00\% & 0.00\% \\
Lla4:16x17b & 0.00\% & 0.00\% & 0.00\% & 0.00\% & 0.00\% & 0.00\% \\
Mist-N:12b & 0.00\% & 0.05\% & 0.00\% & 0.18\% & 0.00\% & 0.00\% \\
Nemo:70b & 0.00\% & 0.14\% & 0.00\% & 0.00\% & 0.00\% & 0.00\% \\
Qw3:235b & \textbf{100\%} & \textbf{100\%} & 0.00\% & 16.96\% & \textbf{100\%} & \textbf{100\%} \\
Qw3:32b & \textbf{100\%} & \textbf{100\%} & \textbf{100\%} & \textbf{100\%} & \textbf{100\%} & \textbf{100\%} \\
\bottomrule
\end{tabular}
}
\end{table}

\begin{table}[ht!]
\centering
\caption{Results for 3D Sierpinski Pyramid.}
\label{tab:3d_pyramid}
\resizebox{\columnwidth}{!}{%
\begin{tabular}{lcccccc}
\toprule
\multirow{2}{*}{\textbf{Model}} & \multicolumn{2}{c}{\textbf{20 pts}} & \multicolumn{2}{c}{\textbf{50 pts}} & \multicolumn{2}{c}{\textbf{100 pts}} \\ \cmidrule(lr){2-3} \cmidrule(lr){4-5} \cmidrule(lr){6-7}
 & \textbf{Ord.} & \textbf{Any} & \textbf{Ord.} & \textbf{Any} & \textbf{Ord.} & \textbf{Any} \\ \midrule
R1:70b & 0.00\% & 0.00\% & 0.00\% & 0.00\% & 0.00\% & 0.00\% \\
Gem3:12b & 0.00\% & 0.20\% & 0.00\% & 0.10\% & 0.00\% (NC) & 0.00\% \\
Gem3:27b & 0.00\% & 0.31\% & 0.00\% & 0.18\% & 0.00\% & 0.00\% \\
OSS:120b & \textbf{100\%} & \textbf{100\%} & 0.00\% & 1.23\% & \textbf{100\%} & \textbf{100\%} \\
OSS:20b & 0.00\% (NC) & 0.00\% & 0.00\% (NC) & 0.00\% & 0.00\% (NC) & 0.00\% \\
Lla3.3:70b & 0.00\% (NC) & 0.59\% & 0.00\% (NC) & 0.00\% & 0.00\% & 0.28\% \\
Lla4:16x17b & 0.00\% & 0.01\% & 0.00\% & 1.87\% & 0.00\% (NC) & 0.00\% \\
Mist-N:12b & 0.00\% & 0.49\% & 0.00\% & 0.00\% & 0.00\% & 0.00\% \\
Nemo:70b & 0.00\% (NC) & 0.00\% & 0.00\% (NC) & 0.00\% & 0.00\% & 2.52\% \\
Qw3:235b & 0.00\% (NC) & 0.00\% & 0.00\% (NC) & 0.00\% & 0.00\% (NC) & 0.00\% \\
Qw3:32b & 0.00\% & 0.01\% & 0.00\% & 0.52\% & 0.00\% (NC) & 0.00\% \\
\bottomrule
\end{tabular}
}
\end{table}

\begin{table}[ht!]
\centering
\caption{Results for 3D Menger Sponge.}
\label{tab:3d_menger}
\resizebox{\columnwidth}{!}{%
\begin{tabular}{lcccccc}
\toprule
\multirow{2}{*}{\textbf{Model}} & \multicolumn{2}{c}{\textbf{20 pts}} & \multicolumn{2}{c}{\textbf{50 pts}} & \multicolumn{2}{c}{\textbf{100 pts}} \\ \cmidrule(lr){2-3} \cmidrule(lr){4-5} \cmidrule(lr){6-7}
 & \textbf{Ord.} & \textbf{Any} & \textbf{Ord.} & \textbf{Any} & \textbf{Ord.} & \textbf{Any} \\ \midrule
R1:70b & 0.00\% & 0.05\% & 0.00\% (NC) & 0.00\% & 0.00\% & 0.05\% \\
Gem3:12b & 0.00\% & 0.05\% & 0.00\% & 0.36\% & 0.00\% & 0.05\% \\
Gem3:27b & 0.00\% & 0.05\% & 0.00\% & 0.05\% & 0.00\% & 0.05\% \\
OSS:120b & 0.00\% & 0.00\% & 0.01\% & 0.16\% & 0.01\% & 0.36\% \\
OSS:20b & 0.00\% & 0.00\% & 0.01\% & 0.16\% & 0.00\% & 0.00\% \\
Lla3.3:70b & 0.00\% & 0.05\% & 0.00\% & 0.04\% & 0.00\% & 0.36\% \\
Lla4:16x17b & 0.00\% & 0.06\% & 0.00\% & 0.16\% & 0.00\% & 0.16\% \\
Mist-N:12b & 0.00\% & 0.03\% & 0.00\% & 0.00\% & 0.00\% & 0.11\% \\
Nemo:70b & 0.00\% (NC) & 0.00\% & 0.00\% & 0.05\% & 0.00\% & 0.01\% \\
Qw3:235b & 0.00\% & 0.05\% & 0.01\% & 0.16\% & 0.00\% (NC) & 0.00\% \\
Qw3:32b & 0.00\% & 0.00\% & 0.00\% & 0.04\% & 0.00\% & 0.14\% \\
\bottomrule
\end{tabular}
}
\end{table}

\subsection{Energy Cost of Symbolic Inference}
\label{subsec:inference_energy}
\begin{figure*}[ht!]
    \centering
    \includegraphics[width=\textwidth]{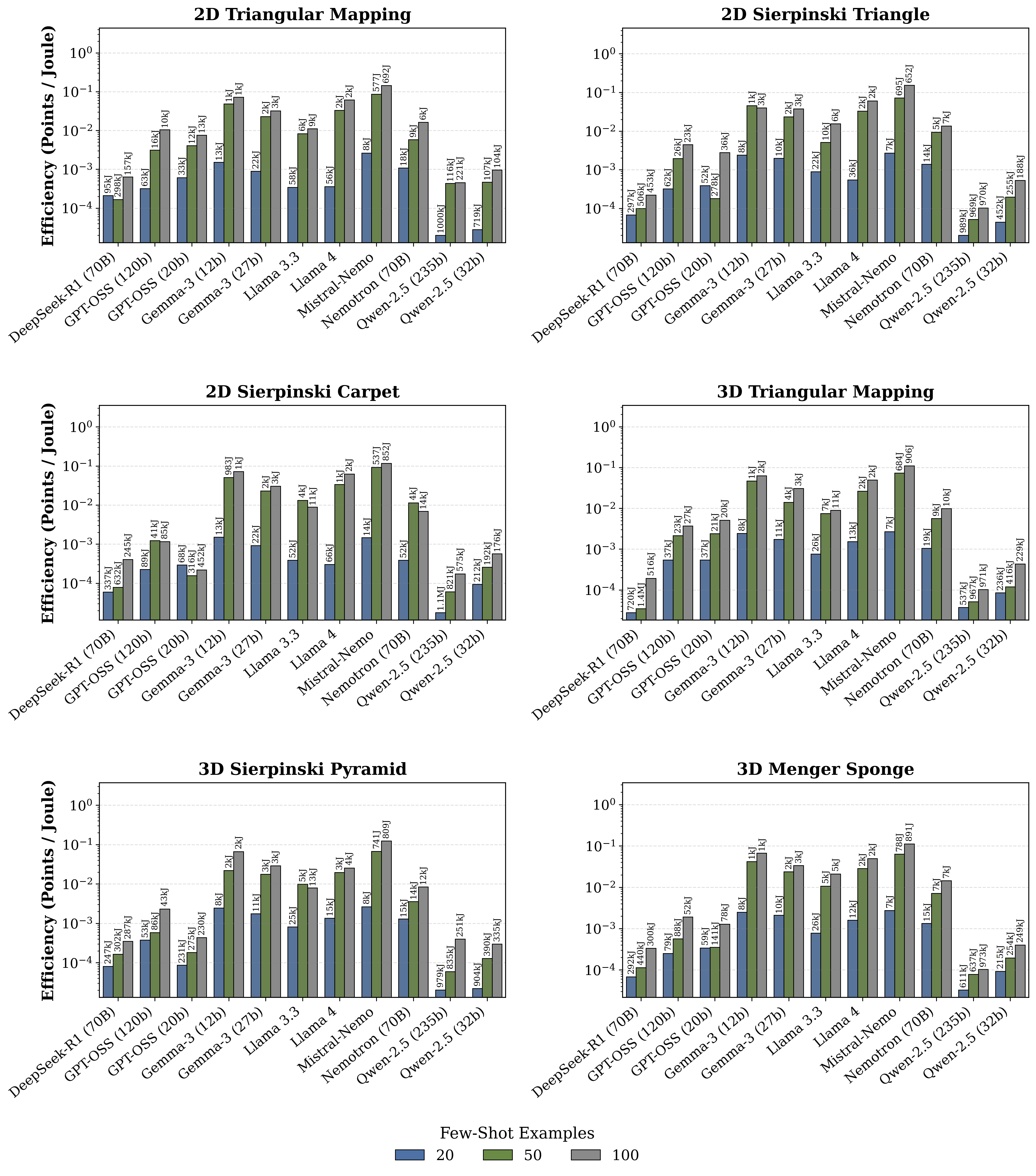}
    \caption{Computational efficiency (Points/Joule) of each open-source model 
    across six spatial domains---three 2D (top) and three 3D (bottom)---evaluated 
    at 20, 50, and 100 few-shot examples. Higher values indicate greater energy 
    efficiency on a logarithmic scale. Bar annotations report the total energy 
    consumed per inference run. Efficiency is defined as the number of correctly 
    mapped points divided by total energy consumed.}
    \label{fig:efficiency_all_domains}
\end{figure*}

We also evaluated the upfront energy cost of running the LLM inference itself. We measured the computational efficiency, defined as Points processed per Joule (Points/Joule), during the equation derivation phase. The hardware resource for this phase was an allocation of 4x NVIDIA A100 (40GB SXM4) GPUs within a DGX node. To ensure clean measurements, all inferences were executed in strict isolation (one model at a time) and energy consumption was recorded using NVIDIA's NVML (nvidia-smi) profiling tools. Figure~\ref{fig:efficiency_all_domains} presents this efficiency metric across all six geometric domains, evaluated at 20, 50, and 100
in-context examples.

\subsubsection{Impact of Model Architecture and Reasoning}
Two distinct efficiency profiles emerge within the open-source ecosystem:

\begin{enumerate}
    \item \textbf{Parameter-Driven Penalties:} Massive models like \texttt{qwen3:235b} exhibit lower points-per-joule efficiency driven by the substantial memory bandwidth required to move 235.1 billion parameters across the 4 GPUs, yielding a high baseline power draw during execution.
    
    \item \textbf{Reasoning-Driven Penalties:} Notably, \texttt{deepseek-r1:70b} often matches or falls below the efficiency of models three times its size. This is attributed to its "Chain-of-Thought" (CoT) mechanism. While standard dense models like \texttt{llama3.3} (70.6B) emit code quickly, DeepSeek-R1 requires significantly more time to "reason" through the problem, imposing a severe energy penalty (fewer points processed per joule) compared to standard models of equivalent parameter count.
\end{enumerate}

\subsubsection{Efficiency Gains from In-Context Learning}
Increasing the number of in-context examples generally improved the energy efficiency. For standard models like \texttt{llama3.3} and \texttt{mistral-nemo}, moving from 20 to 50 or 100 shots resulted in a noticeable increase in efficiency. Qualitative analysis suggests that with fewer examples (20 shots), models are prone to generating verbose code that fails to compile or "hallucinating" lengthy explanations. Providing a richer context constrains the generation space, guiding the model to produce concise, correct solutions faster, thereby minimizing GPU active time per valid point. For \texttt{deepseek-r1}, it occasionally shows an efficiency drop at 50 shots, possibly due to an increased activation of its internal reasoning traces before converging efficiently at 100 shots.

\subsection{Performance and Energy in Block-Level Execution}
\label{subsec:execution_energy}
Once the LLM successfully infers the correct analytical GPU thread 
function, this one-time upfront inference cost allows doing the actual 
GPU work by just evaluating the inferred map with all threads, not 
needing to infer again. To quantify this, the analytical expressions 
derived by the LLMs were tested with a lightweight dummy CUDA kernel 
that performs an atomic increment (\texttt{+1}) on each mapped memory 
address, simulating a representative memory-bound workload without 
introducing algorithm-specific overhead. This kernel was integrated 
into a CUDA block-level mapping kernel which is the state of the art 
way of applying efficient GPU thread mapping~\cite{ref1,ref2,ref6,ref7}.

For these performance and power profiling tests, the kernels were 
executed on a single, isolated NVIDIA A100 (40GB SXM4) GPU within the 
DGX node, using NVML for precise energy polling. We report the gross energy draw, which includes both the 
dynamic calculation energy and the hardware's idle baseline power. Each 
table includes two reference baselines: the \textit{Bounding Box} (BB), 
representing the naive mapping strategy, and the \textit{Paper} entry, 
corresponding to the analytically derived mapping from the established 
literature, which serves as the human-effort gold standard against 
which LLM-generated solutions are compared.

\textbf{Dense Geometries:} As shown in Table~\ref{tab:block_dense}, 
the use of optimal $O(1)$ cost analytical functions generated by top 
models eliminates the resource waste inherent in the Bounding Box (BB) 
method without sacrificing performance. In the \textbf{3D Pyramid} 
domain, the BB approach wastes approximately 83\% of the launched 
blocks, resulting in an execution time of 2530.65 ms and consuming over 
282 Joules. In contrast, the optimal inferred kernels reduce this time 
to 3.84 ms and consume less than 1 Joule. However, the results 
highlight the penalty of inefficient code generation: while 
\texttt{OSS:120b} (Stage 20) correctly solved the mapping logic, 
it implemented a Linear Search algorithm ($O(N^{1/3})$)---so named 
because it iterates linearly over candidate integer values of the depth 
parameter until the boundary condition is satisfied, despite the 
closed-form involving a cube root---rather than a direct analytical 
solution, degrading performance to 117.03 ms and consuming 
22.25 J---over $24\times$ more energy than the optimal solution.

\textbf{Fractal Geometries:}
The impact of exact analytical mapping is most pronounced in fractal 
domains (Table~\ref{tab:block_fractal}), where the sparsity of the 
geometry renders the BB approach unfeasible for large domains. For the 
\textbf{3D Sierpinski} structure, the BB method requires traversing a 
dense enclosing cube, launching over 8 billion blocks to identify only 
1.9 million valid ones of the actual fractal geometry which is not 
dense. This results in an estimated execution time of nearly 16 seconds (projected from a smaller scale due to extreme physical timeout overhead) 
and an energy cost of $\sim 1591$ J. The bitwise operations inferred by 
the top-performing LLMs allow for direct indexing of valid blocks, 
reducing execution time to 3.30 ms and energy consumption to just 0.55 
Joules. This represents an effective speedup of $\sim 4833\times$ and 
an energy reduction of $2890\times$, instantly amortizing the initial 
LLM inference energy cost on the very first execution.

\subsection{Discussion: Parameter Scaling vs. Reasoning}
Our findings challenge the traditional scaling laws in the context of exact symbolic derivation. While large dense models like \texttt{qw3:235b} exhibit high memory bandwidth and energy demands, they do not inherently guarantee success in complex domains like the 3D Sierpinski Pyramid. Conversely, reasoning-focused architectures (e.g., \texttt{R1:70b}) leverage latent Chain-of-Thought processing to correctly identify recursive mapping patterns that defeat larger parameter-dense counterparts.

Ultimately, the boundaries observed on the 3D Menger Sponge outline a current ``Menger Limit'' for the open ecosystem. This domain forces the LLM to simultaneously track cubic index progressions and non-trivial 3D void removals, marking the absolute cutting edge of open-source symbolic inference. We note that this represents the current state of open-weights training maturity rather than a fundamental limitation in the LLM paradigm. As open models continue to adopt more advanced techniques—such as sophisticated Mixture-of-Experts (MoE) routing and deeper reinforcement learning—we anticipate this current frontier will soon be surpassed. From a practical standpoint, this implies that researchers do not necessarily need massive 200B+ parameter models requiring multi-GPU clusters; highly optimized reasoning models can achieve state-of-the-art symbolic derivation on a single node.

\subsection{Applicability, Limitations, and Framework Extensibility}
\label{subsec:limitations}
The proposed framework is specifically applicable to computational domains that exhibit deterministic mathematical patterns, such as arithmetic progressions (dense geometries) or recursive self-similarity (fractals). It is not designed for completely unstructured meshes where coordinate mapping depends on arbitrary data points without an underlying mathematical law.

While a human expert could manually interact with an LLM to derive an equation for a singular geometry, the primary value of this framework lies in the complete automation of the pipeline, removing human analytical effort from the compilation or execution stages of HPC workloads.

Furthermore, the system is fundamentally modular and model-agnostic. As new, more capable open-weight architectures are released, they can be deployed within the pipeline without modifying the integration logic. Regarding the in-context learning process, the structured prompt detailed in Appendix A is empirically robust in enforcing strict algorithmic generation. We do not assert that it constitutes the absolute theoretical optimum of prompt engineering; rather, it establishes a reliable, reproducible baseline necessary to evaluate and expose the current mathematical reasoning limits of existing neural architectures.

\begin{table*}[ht!]
    \centering
    \caption{CUDA Block-Level Performance and Energy in Dense Geometries ($N=500\times 10^6$).}
    \label{tab:block_dense}
    \setlength{\tabcolsep}{5pt}
    \begin{tabular}{llrrrrl}
        \toprule
        \textbf{Domain} & \textbf{Model (Stage)} & \textbf{Time (ms)} & \textbf{Total Blocks} & \textbf{Wasted} & \textbf{Energy (J)} & \textbf{Logic (Complexity)} \\
        \midrule
        \multirow{10}{*}{\textbf{2D Triangular}} 
        & Bounding Box (Baseline) & 747.45 & 3,912,484 & 1,959,359 & 83.27 & If ($O(1)$) \\
        & Paper (Navarro 2014) & 1.46 & 1,953,125 & 0 & 0.44 & Analytical ($O(1)$) \\
        & R1:70b (Stage 20) & 1.46 & 1,953,125 & 0 & 0.45 & Analytical ($O(1)$) \\
        & R1:70b (Stage 50) & 1.46 & 1,953,125 & 0 & 0.45 & Analytical ($O(1)$) \\
        & OSS:120b (All Stages) & 1.46 & 1,953,125 & 0 & 0.45 & Analytical ($O(1)$) \\
        & Lla3.3:70b / Nemo:70b & 1.46 & 1,953,125 & 0 & 0.45 & Analytical ($O(1)$) \\
        & R1:70b (Stage 100) & 1.97 & 1,953,125 & 0 & 0.70 & Sqrt+Loops ($O(1)$) \\
        & OSS:20b (Stage 50) & 1.51 & 1,953,125 & 0 & 0.51 & Approx+If ($O(1)$) \\
        & OSS:20b (Stage 100) & 1.51 & 1,953,125 & 0 & 0.51 & Approx+If ($O(1)$) \\
        & Qw3:32b (Stage 50) & 14.86 & 1,953,125 & 0 & 3.21 & BinSearch ($O(\log N)$) \\
        \midrule
        \multirow{9}{*}{\textbf{3D Pyramid}} 
        & Bounding Box (Baseline) & 2530.65 & 12,008,989 & 10,055,864 & 282.67 & If ($O(1)$) \\
        & Paper (Navarro 2016) & 3.84 & 1,953,125 & 0 & 0.92 & Analytical ($O(1)$) \\
        & R1:70b (Stage 50) & 6.21 & 1,953,125 & 0 & 1.44 & Cbrt+Loop ($O(1)$) \\
        & Qw3:32b (All Stages) & 6.21 & 1,953,125 & 0 & 1.44 & Cbrt+Loop ($O(1)$) \\
        & OSS:120b (Stage 100) & 29.31 & 1,953,125 & 0 & 5.99 & BinSearch ($O(\log N)$) \\
        & Qw3:235b (Stage 20) & 29.31 & 1,953,125 & 0 & 5.99 & BinSearch ($O(\log N)$) \\
        & OSS:120b (Stage 50) & 51.57 & 1,953,125 & 0 & 9.12 & BinSearch+Lin ($O(N^{1/3})$) \\
        & OSS:120b (Stage 20) & 117.03 & 1,953,125 & 0 & 22.25 & Linear ($O(N^{1/3})$) \\
        \bottomrule
    \end{tabular}
\end{table*}

\begin{table*}[ht!]
    \centering
    \caption{CUDA Block-Level Performance and Energy in Fractal Geometries ($N=500\times 10^6$).}
    \label{tab:block_fractal}
    \setlength{\tabcolsep}{6pt}
    \begin{tabular}{llrrrrl}
        \toprule
        \textbf{Domain} & \textbf{Model (Stage)} & \textbf{Time (ms)} & \textbf{Total Blocks} & \textbf{Wasted} & \textbf{Energy (J)} & \textbf{Logic (Complexity)} \\
        \midrule
        \multirow{3}{*}{\textbf{2D Sierpinski}} 
        & Bounding Box (Baseline) & 65.78 & 88,736,400 & 86,783,275 & 6.73 & If ($O(1)$) \\
        & Paper (Reference) & 8.62 & 1,953,125 & 0 & 1.39 & Bitwise ($O(\log N)$) \\
        & OSS:120b (Stage 20) & 8.62 & 1,953,125 & 0 & 1.39 & Bitwise ($O(\log N)$) \\
        \midrule
        \multirow{3}{*}{\textbf{3D Sierpinski}} 
        & Bounding Box (Baseline)* & 15,949.00 & 8,000,000,000 & 7,998,046,875 & 1,591.71 & If ($O(1)$) \\
        & Paper (Reference) & 3.30 & 1,953,125 & 0 & 0.55 & Bitwise ($O(\log N)$) \\
        & R1:70b (Stage 100) & 3.30 & 1,953,125 & 0 & 0.56 & Bitwise ($O(\log N)$) \\
        \bottomrule
        \multicolumn{7}{l}{\footnotesize * Baseline projected from $N=5\times 10^6$ due to physical timeout on extreme waste overhead.}
    \end{tabular}
\end{table*}

\section{Code and Data Availability}
\label{sec:code_availability}
The source code, generated domain datasets, and CUDA evaluation kernels associated with this manuscript are publicly available to ensure full reproducibility. The repository can be accessed at \url{https://github.com/aspiadevs/llm-gpu-thread-mapping}.

\section{Conclusions and Future Work}
\label{sec:conclusions}
This research demonstrates the viability of utilizing state-of-the-art open-source LLMs to automatically infer exact symbolic equations for GPU thread mapping, serving as an alternative to manual human mathematical derivation. More in detail, three findings are worth highlighting:
\begin{enumerate}
    \item \textbf{Open-Source LLM Viability:} Open-weights LLMs can infer efficient GPU thread maps for 2D/3D regular and fractal geometries. In comparison, a traditional Symbolic Regression (SR) approach was unable to deliver accurate symbolic functions for GPU thread mapping. 
    \item \textbf{The Energy Trade-off:} Automating this derivation locally exposes significant energy disparities. While generating the code incurs a high one-time upfront energy cost---especially when utilizing reasoning-heavy models like DeepSeek-R1---this investment is instantly amortized. Once integrated, the inferred GPU maps eliminate the unnecessary blocks, saving orders of magnitude in execution time and energy (up to $4833\times$ faster and $2890\times$ more energy efficient) compared to the baseline).
    \item \textbf{The Menger Frontier:} While open models excel in many spatial domains, highly recursive 3D fractal structures (such as the Menger Sponge) represent the current reasoning ceiling for the open ecosystem. This establishes a clear, measurable benchmark for future open-weights architectures to overcome as their mathematical reasoning capabilities evolve.
\end{enumerate}

\textbf{Future Work:}
Future research will focus on mitigating the upfront energy penalty of LLM inference by exploring the targeted fine-tuning of smaller, highly efficient open-weights models (e.g., sub-10B parameters) specifically specialized for discrete spatial reasoning. This approach aims to deliver the exactness of massive reasoning models at a fraction of the initial computational cost. Additionally, we plan to extend this automated mapping framework to encompass more heterogeneous HPC topologies, such as unstructured meshes and adaptive refinement grids, while continuously monitoring the evolution of open-source architectures to verify when the 3D recursive fractal boundary is definitively breached.

\section*{Acknowledgment}
This work was supported by Universidad Austral de Chile and ANID FONDECYT grants \#1221357, \#1261152, the Temporal research group and the Patagón supercomputer of Universidad Austral de Chile (FONDEQUIP EQM180042).

\appendices

\section{Detailed Prompt Specification}
\label{app:prompt}

To guide the models in equation inference, a detailed prompt was designed that specifies the expert role, the inference task, and the output requirements. The data points for each test case were appended to the following template:

\begin{lstlisting}[breaklines=true, basicstyle=\ttfamily\small, frame=single, backgroundcolor=\color{gray!10}]
<ROLE>
Act as an expert in mathematics and cryptography, specializing in the reverse engineering of algorithms and the identification of complex patterns in multidimensional spaces. Your goal is SOLELY to generate the Python code requested.
</ROLE>

<TASK>
Analyze the mapping data in the <CONTEXT> to find the underlying mathematical algorithm.
Then, generate the complete source code for a single Python function that implements this general algorithm.
</TASK>

<CONTEXT>
# Mapping Data
__MAPPING_DATA_HERE__
</CONTEXT>

<RULES>
- Function name must be exactly `map_to_coordinates(n)`.
- Input: `n` (non-negative integer).
- Output: tuple of integers representing coordinates.
- Each integer within the returned coordinate tuple must be greater than or equal to 0.
- Validate input `n` (non-negative integer), raise `ValueError` if invalid.
- **CRITICAL ALGORITHM CONSTRAINT:** The function MUST implement a general mathematical algorithm that works for ANY non-negative integer 'n', not just the examples provided.
- **DO NOT use hardcoded values, lookup tables, or long `if/elif` chains based on ranges of 'n' (e.g., `if n == 1:`, `if n < 10:`, `if 10 <= n <= 20:` are forbidden).**
- **CRITICAL OUTPUT CONSTRAINT:** Your response MUST contain ONLY the Python code block for the function.
- **DO NOT include ANY introductory text, explanations, reasoning, thought processes, or comments (including docstrings or # comments) inside or outside the code block.**
- Do NOT include an `if __name__ == "__main__":` block.
</RULES>

<RESPONSE>
\end{lstlisting}

\begin{IEEEbiography}[{\includegraphics[width=1in,height=1.25in,clip,keepaspectratio]{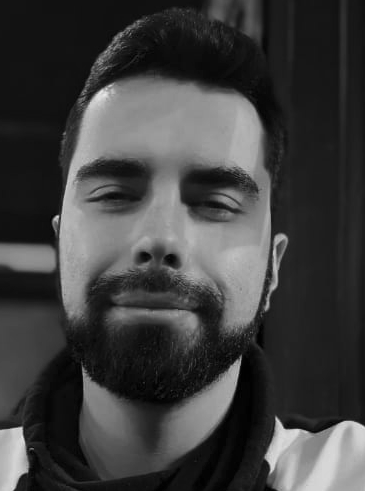}}]{Jose Maureira}
received the M.Sc. degree in Computer Science from the Universidad Austral de Chile in 2026, where he is currently pursuing the Ph.D. degree in Engineering Sciences. He is a Research Assistant with the Temporal research lab, and his research interests include high-performance computing, large language models, computer vision, and the generation of synthetic data.
\end{IEEEbiography}
\vspace{-4.0\baselineskip}

\begin{IEEEbiography}[{\includegraphics[width=1in,height=1.25in,clip,keepaspectratio]{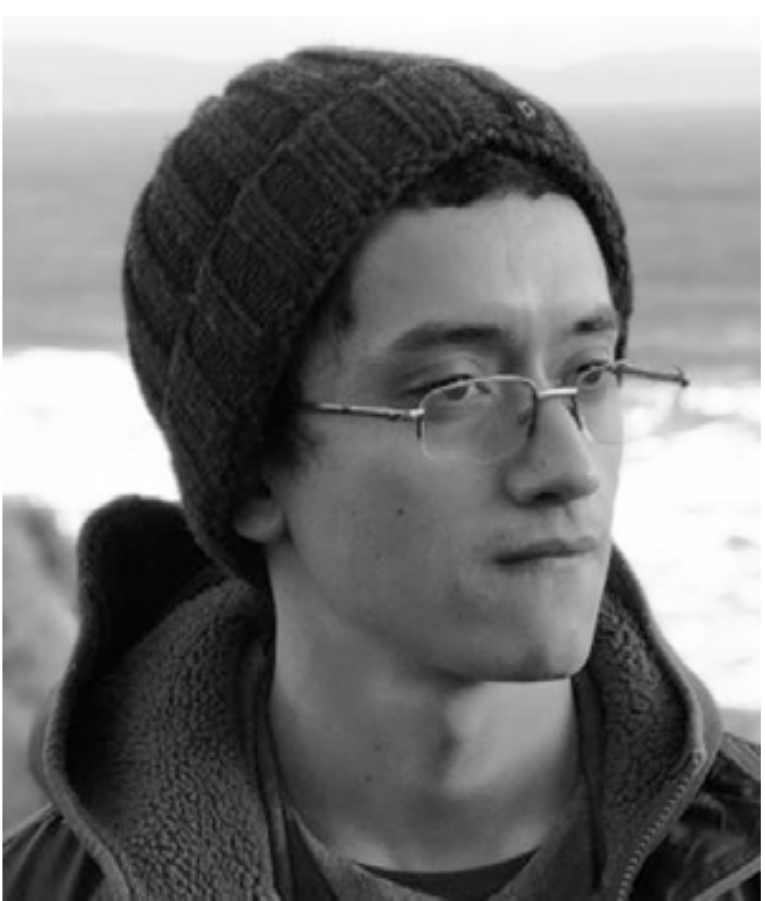}}]{Crist\'obal A. Navarro}
has a Ph.D. degree in computer science from the University of Chile (2015). Currently, he is an associate professor at the Universidad Austral de Chile and leads the \textit{Temporal} research lab as well as the Patagón Supercomputer project. Today, his research interests include GPU computing, computer graphics, and computational physics.
\end{IEEEbiography}
\vspace{-4.0\baselineskip}

\begin{IEEEbiography}[{\includegraphics[width=1in,height=1.25in,clip,keepaspectratio]{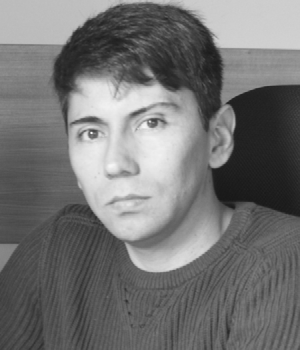}}]{Hector Ferrada}
received his Ph.D. degree in Computer Science from the University of Chile in 2016, focusing on his research in the design and analysis of algorithms for compact data structures. In 2016-2017, he conducted postdoctoral research in the Genome Scaling Algorithms Group at the University of Helsinki, Finland, in collaboration with Dr. Veli Mäkinen. Currently, he mainly teaches courses related to his research interests in algorithms and data structures. 
\end{IEEEbiography}
\vspace{-4.0\baselineskip}

\begin{IEEEbiography}[{\includegraphics[width=1in,height=1.25in,clip,keepaspectratio]{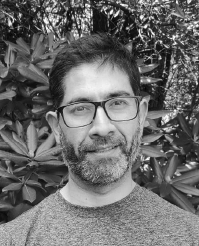}}]{Luis Veas-Castillo}
received the Ph.D. degree in Engineering Sciences from the Universidad de Santiago de Chile in 2022. His research interests include parallel and distributed systems, computational simulation, large-scale databases, and applied artificial intelligence. He has extensive experience leading interdisciplinary R\&D projects in collaboration with Chilean institutions and research centers. He is currently driving applied research initiatives at the Universidad Austral de Chile, integrating engineering with healthcare and environmental sciences.
\end{IEEEbiography}

\end{document}